\PassOptionsToPackage{english}{babel}
\documentclass[reprint,aps,pra,twocolumn,showpacs,amssymb,amsfonts,superscriptaddress,longbibliography]{revtex4-1}  
\usepackage[english]{babel}
\usepackage{graphicx}  
\usepackage{dcolumn}   
\usepackage{bm}        
\usepackage{bbm}        
\usepackage{amssymb}   
\usepackage{amsmath}
\usepackage{mathtools}
\usepackage[utf8]{inputenc}
\usepackage{siunitx}

\usepackage{changes}

\definecolor{myurlcolor}{rgb}{0,0,0.7}
\definecolor{myrefcolor}{rgb}{0.8,0,0}
\usepackage{hyperref}
\hypersetup{colorlinks, linkcolor=myrefcolor,
citecolor=myurlcolor, urlcolor=myurlcolor}

\usepackage{cleveref}
\crefrangeformat{equation}{#3(#1 #4--#5 #2)#6}

\usepackage{comment}
\usepackage{soul} 
\newcommand{\ignore}[1]{}
\usepackage{enumitem}

\newcommand{\eg}{\textit{e.g. }}				
\newcommand{\ie}{\textit{i.e. }}				
\newcommand{\var}{\ensuremath{\, \text{Var}}}
\newcommand{\cov}{\ensuremath{\, \text{Cov}}}

\newcommand{\vect}[1]{{\bm{#1}}}			
\newcommand{\avg}[1]{\left\langle #1 \right\rangle}		

\newcommand{\bra}[1]{\ensuremath{\langle#1|}}
\newcommand{\ket}[1]{\ensuremath{|#1\rangle}}

\newcommand{\abs}[1]{\ensuremath{\left\vert #1\right\vert}}

\newcommand{\SuA}{S_{\vec{u}}^A}
\newcommand{\su}{s_{\vec{u}}}
\newcommand{\sui}{s_{\vec{u}}^{(i)}}

\newcommand{\SuB}{S_{\vec{u}}^B}

\newcommand{\Su}{S_{\vec{u}}}
\newcommand{\SuX}{S_{\vec{u}}^I}
\newcommand{\SuY}{S_{\vec{u}}^J}
\newcommand{\SvY}{S_{\vec{v}}^J}
\newcommand{\PX}{\Pi^{I}}
\newcommand{\PY}{\Pi^{J}}

\newcommand{\sv}{s_{\vec{v}}}
\newcommand{\svi}{s_{\vec{v}}^{(i)}}
\newcommand{\svj}{s_{\vec{v}}^{(j)}}

\newcommand{\PA}{\Pi^{A}}

\newcommand{\PXi}{\Pi^{I,(i)}}
\newcommand{\PYi}{\Pi^{J,(i)}}
\newcommand{\PYj}{\Pi^{J,(j)}}

\newcommand{\PB}{\Pi^{B}}

\begin{document}
\selectlanguage{english}

\title{Relating spin squeezing to multipartite entanglement criteria for particles and modes}
\author{Matteo Fadel}
\email{matteo.fadel@unibas.ch} 
\affiliation{Department of Physics, University of Basel, Klingelbergstrasse 82, 4056 Basel, Switzerland} 
\author{Manuel Gessner}
\email{manuel.gessner@ens.fr}
\affiliation{Laboratoire Kastler Brossel, ENS-Universit\'{e} PSL, CNRS, Sorbonne Universit\'{e}, Coll\`{e}ge de France, 24 Rue Lhomond, 75005, Paris, France}

\date{\today}

\begin{abstract}
Entanglement witnesses based on first and second moments exist in the form of spin-squeezing criteria for the detection of particle entanglement from collective measurements, and in form of modified uncertainty relations for the detection of mode entanglement or steering. By revealing a correspondence between them, we show that metrologically useful spin squeezing reveals multimode entanglement in symmetric spin states that are distributed into addressable modes. We further derive tight state-independent multipartite entanglement bounds on the spin-squeezing coefficient and point out their connection to widely-used entanglement criteria that depend on the state's polarization. Our results are relevant for state-of-the-art experiments where symmetric entangled states are distributed into a number of addressable modes, such as split spin-squeezed Bose-Einstein condensates.
\end{abstract}

\maketitle

\section{Introduction}
Quantum entanglement describes non-classical correlations of multipartite quantum systems~\cite{Horodecki}. It can appear between the parties' internal (\eg spin) degree of freedom (dof), or between their external (\eg spatial modes) dof. One usually refers to these two cases as particle or mode entanglement, respectively.  Apart from its fundamental interest, entanglement is a key resource in quantum information science and quantum technologies~\cite{Bollinger,PS09}. This is evidenced, for instance, in the context of quantum sensing and metrology, where quantitative relations between metrological sensitivity and the number of entangled particles in Ramsey interferometers exist~\cite{HyllusToth}. In atomic ensembles, entangled multipartite quantum states with the potential to enhance interferometric measurements can be prepared by controlling the interactions between particles, which is a well-established technique in today's experiments~\cite{RMP}.

\begin{figure}
  \centering
\includegraphics[width=0.49\textwidth]{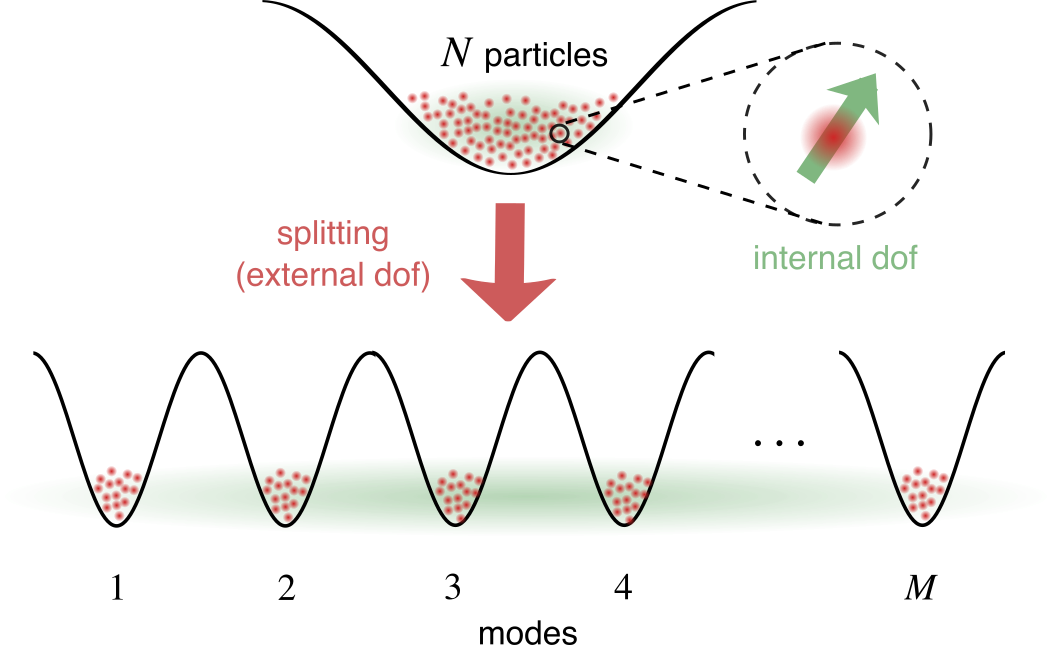}
  \caption{By spatially splitting an entangled ensemble of $N$ identical particles into $M$ external modes, we generate entanglement between addressable modes.}
    \label{fig:scheme}
\end{figure}

Most experiments on ultracold atomic ensembles focus on quantum states where particles share the same external mode, and can thus only be addressed and measured collectively. In recent years, new technologies, such as quantum gas microscopes~\cite{BakrQuantumGasMicroscope}, optical tweezer traps~\cite{Browaeys,Lukin}, and split Bose-Einstein condensates (BECs)~\cite{FadelSplitBEC,KunkelSplitBEC,LangeSplitBEC} have enabled the investigation of spatially distributed, entangled atomic ensembles, see Fig.~\ref{fig:scheme}. In such systems, on top of the entanglement among the particles, we can study the entanglement of spatially separated modes~\cite{Oudot,Jing}. On the one hand, this is interesting for practical applications such as spatially-resolved metrology~\cite{HumphreysPRL2013,ProctorPRL2018,GePRL2018,GessnerPRL2018}, optical clocks \cite{Kajtoch18}, and quantum information tasks~\cite{Nielsen2000}. On the other hand, it allows to investigate fundamental concepts such as the extraction of entanglement from a system of indistinguishable particles \cite{Killoran14,LoFranco18,Morris19,Barros19,Sun20}.

Particle entanglement can be revealed experimentally through spin-squeezing coefficients~\cite{Wineland,SorensenNature,SorensenMolmer,Ma,GuehneToth,Hyllus1,Hyllus2}. Among several methods to quantify spin squeezing~\cite{Ma}, the Wineland \textit{et al.} spin-squeezing coefficient~\cite{Wineland} has the advantage of establishing a link between the entanglement detected in a quantum state and its quantum gain for interferometric measurements, thereby detecting metrologically useful entanglement~\cite{PS09}. Moreover, this spin-squeezing coefficient expresses a sensitivity gain that can be reached by a simple parameter estimation protocol after sufficiently many experimental repetitions \cite{Wineland}. Being a function of averages and variances of linear operators only, spin-squeezing coefficients are particularly suitable to detect multiparticle entanglement of spin states that can be approximated as Gaussian quantum states.

While particle entanglement can be detected with collective measurements, standard methods to reveal mode entanglement require local measurements on each mode. Criteria based on variances and mean values can be found in the form of modified uncertainty relations that hold for arbitrary separable states, but can be violated through entanglement~\cite{DGCZ,Raymer,Giovannetti,Werner,HyllusEisert}. These approaches are powerful tools to detect entanglement in arbitrary-dimensional systems with high flexibility. They allow us to study entanglement between specific partitions of a composite system, thereby providing precise microscopic information about which subsystems share quantum correlations~\cite{GessnerQuantum,QinNPJQI2019}. These criteria exist for both discrete and continuous variables, and they can be extended to study the stronger class of quantum correlations known as steering that is at the heart of the Einstein-Podolsky-Rosen (EPR) paradox~\cite{Reid1989,ReidRMP}.

In this work, we show that under conditions that hold for a wide range of systems, the uncertainty-type mode-entanglement criteria coincide with the Wineland \textit{et al.} spin-squeezing coefficient, which detects particle entanglement and quantifies the metrological quantum gain. This allows us to establish a direct relation between the detected entanglement of particles and modes, as well as the sensitivity of spin states. While this does not replace the need for multimode entanglement witnesses, our results reveal the required level of spin squeezing for the generation of multimode entanglement by distributing the spins into addressable modes, e.g., by splitting a BEC into individually addressable ensembles. By linking these quantities to the spin-squeezing coefficient, we further relate mode entanglement and EPR steering to the quantum advantage in metrology measurements. Finally, we improve the best known bounds on the number of entangled particles that can be identified from the spin-squeezing coefficient without knowledge of the average polarization, and we clarify the connection to the spin-squeezing multipartite entanglement criterion by S\o{}rensen and M\o{}lmer~\cite{SorensenMolmer}.

\section{Mode vs particle entanglement}
A collection of systems (labeled $1,...,\Xi$) is entangled if their quantum state cannot be written as
\begin{equation}\label{entdef}
    \rho = \sum_\gamma p_\gamma \rho_\gamma^{(1)} \otimes \cdots\otimes \rho_\gamma^{(\Xi)} \;,
\end{equation}
where $\sum_\gamma p_\gamma=1$ is a probability distribution, and $\rho_\gamma^{(i)}$ are density matrices for system $i$. The local systems may refer either to the $\Xi=N$ particles or to the $\Xi=M$ modes that they occupy, giving rise to particle or mode entanglement, respectively. In practice, determining whether a given quantum state allows for a decomposition of the form of Eq.~\eqref{entdef} is an extremely hard task. One therefore relies on entanglement witnesses or, more generally, necessary conditions that any separable state must satisfy~\cite{GuehneToth,HuberReview}. A violation of these criteria then represents a witness for entanglement.

\subsection{Uncertainty-based mode-entanglement criterion}
Criteria based on first and second moments of linear observables are powerful tools to detect entanglement~\cite{DGCZ,Raymer,Hofmann,GuehneToth} and steering~\cite{Reid1989,ReidRMP} in arbitrary-dimensional systems with high flexibility. An important class of these criteria take on the form of Heisenberg-Robertson-type uncertainty relations, with a modified lower bound on the variances that can be violated by entangled states. The most general formulation of these criteria for bipartite systems was given by Giovannetti \textit{et al.} in Ref.~\cite{Giovannetti}, where it was furthermore shown that (nonlinear) product criteria are generally more powerful than (linear) sum criteria (see also \cite{HyllusEisert,ReidRMP}). In the context of atomic spin ensembles it is convenient to express these criteria in terms of collective spin observables. For the case of $N$ spins distributed into $M=2$ modes, labeled as $A, B$, these take the form $\vect{S}^A=\sum_{i\in A} \vect{s}^{(i)}/2$, $\vect{S}^B=\sum_{i\in B} \vect{s}^{(i)}/2$, where $\vect{s}^{(i)}$ is the spin for particle $i$. The criterion expresses that all mode-separable states satisfy
\begin{equation}\label{Giova}
\mathcal{G}^2 := \dfrac{4 \var\left[S_z^A + S_z^B \right] \var\left[S_y^A - S_y^B \right] }{\left( \vert\langle S_x^A\rangle\vert + \vert\langle S_x^B\rangle\vert \right)^2} \geq 1 \;.
\end{equation}
The choice of observables (\eg local spin components) can further be optimized to identify the most sensitive entanglement criterion for a given quantum state. These criteria can be generalized to study entanglement in specific multipartitions (with precise microscopic information about which subsystems share quantum correlations)~\cite{GessnerQuantum} as well as full inseparability~\cite{VanLoock,Toscano,ReidTeh,RiedTehSpin}, \ie the violation of these bounds in all possible partitions.

\subsection{Spin squeezing particle-entanglement criterion}
In the case of a large number of spins, it becomes challenging to address each particle individually. Nevertheless, particle entanglement among the individual spins can be detected from collective measurements of the spin components through spin-squeezing criteria. For all fully separable spin$-1/2$ states it holds that
\begin{equation}\label{Wine}
\xi^2 := \dfrac{N \var\left[S_z\right]}{\vert\langle S_x\rangle\vert^2} \geq 1 \;,
\end{equation}
where $\xi^2$ is the Wineland \textit{et al.} spin-squeezing coefficient~\cite{Wineland}. States with $\xi^2<1$ are characterized by a variance of the collective spin operator that is smaller than that of a coherent spin state, while at the same time being strongly polarized along the $S_x$ direction.

The spin-squeezing coefficient can be considered as a simple Gaussian approximation of the full metrological sensitivity that can be extracted from the quantum state~\cite{PS09,RMP,GessnerPRL2019}. For this reason, states with $\xi^2<1$ can achieve a quantum enhancement beyond the standard quantum limit in metrology measurements~\cite{Wineland} and the entanglement revealed by this condition is metrologically useful. This approach can be extended to fluctuating particle numbers~\cite{Hyllus1}, multipartite entanglement~\cite{HyllusToth,GessnerPRA2017}, Bell nonlocality~\cite{Frowis}, and to analyze the multimode entanglement structure in addressable systems of arbitrary dimension~\cite{GessnerQuantum,QinNPJQI2019}.

\section{Equivalence of mode and particle entanglement: Two-mode case}\label{sec:symmetry}
In the following we prove that, if the state of the system is symmetric under the exchange of particle labels and of modes, the two criteria~(\ref{Giova}) and~(\ref{Wine}) are equivalent, namely that
\begin{equation}\label{GWequi}
    \mathcal{G}^2 = \xi^2 \;.
\end{equation}
To show this, imagine a system of $i=1,\dots,N$ particles with an internal (spin) and an external (mode) dof. We associate to each particle the operator $\sui \PXi$, where $\sui$ is the spin operator along direction $\vec{u}$, and $\PXi$ is the projection operator of the external dof onto one of the $M=2$ modes labeled as $I=1\equiv A$ and $I=2\equiv B$.
Let us now assume the following properties valid for all $i,j=1,\dots,N$, and $I,J=1,\dots,M \geq 2$:
\begin{enumerate}[label=(\roman*)]
    \item \textbf{dof's factorize}: there are no correlations between the spin and the spatial dof, \eg $\avg{\sui\PXi}=\avg{\sui}\avg{\PXi}$, $\avg{\sui\PXi\svj\PYj}=\avg{\sui\svj}\avg{\PXi\PYj}$;
    \item \textbf{particle symmetry}: the state is invariant under permutations of the particle labels, \eg $\avg{\sui}=\avg{\su^{(1)}}$, $\avg{\sui\svj}=\avg{\su^{(1)}\sv^{(2)}}$ and $\avg{\PXi}=\avg{\Pi^{I,(1)}}$, $\avg{\PXi\PYj}=\avg{\Pi^{I,(1)}\Pi^{J,(2)}}$, for $i\neq j$;
    \item \textbf{symmetric splitting}: a) there is equal probability for a particle to be found in any of the modes, \ie $\avg{\PXi}=1/M$, and b) these probabilities are independent, \ie $\avg{\PXi\PYj}=\avg{\PXi}\avg{\PYj}$.
\end{enumerate}

Let us mention that these assumptions are relevant for a number of experimental systems. For example, they apply to an ensemble of identical atoms distributed symmetrically in a set of external modes, as in Refs.~\cite{FadelSplitBEC,KunkelSplitBEC,LangeSplitBEC}.

Exploiting assumptions (i) and (ii), we can now compute expectation values of collective spin observables as
\begin{align}
\avg{ \SuX } &= \sum_{i=1}^N \avg{\sui \PXi}\nonumber\\
&= \avg{\PX} N \avg{\su} \label{evS}
\end{align}
Note that here, and in the following, we use the shorthand notation $\avg{\su^{(1)}}=\avg{\su}$ and $\avg{\Pi^{I,(1)}}=\avg{\PX}$.

Similarly, we obtain that correlators take the form (see Appendix~\ref{app:A1corr} for details)
\begin{align}
    \avg{ \SuX \SvY } = & \delta_{I,J} \avg{\PX} N \avg{\su^{(1)} \sv^{(2)}} + \nonumber\\
    & + \avg{\PX} \avg{\PY} N(N-1) \avg{\su^{(1)} \sv^{(2)}}  \;. \label{evSS}
\end{align}

To prove now the relation Eq.~\eqref{GWequi}, we use Eqs.~(\ref{evS}) and~(\ref{evSS}) to rewrite the variance appearing in Eq.~\eqref{Giova} as (see the Appendix~\ref{app:A2VarY} for a detailed derivation)
\begin{equation}\label{VyG}
    \var\left[S_y^A - S_y^B \right] = 2 \langle\PA\rangle \dfrac{N}{4} = \dfrac{N}{4} \;\quad\text{using (iii)}.
\end{equation}
For the other variance in Eq.~\eqref{Giova} we can simply write
\begin{equation}\label{VzG}
    \var\left[S_z^A + S_z^B \right] = \var\left[S_z\right] \;.
\end{equation}
The same holds also for the denominator, which can be written as
\begin{equation}\label{xG}
    \left( \vert\langle S_x^A\rangle\vert + \vert\langle S_x^B\rangle\vert \right)^2 = \vert\langle S_x\rangle\vert^2 \;,
\end{equation}
since symmetry implies that $\langle S_x^A\rangle$ and $\langle S_x^B\rangle$ have the same sign. It is now straightforward to combine the results of Eqs.~(\ref{VyG}), (\ref{VzG}), and~(\ref{xG}) to see that, under the assumptions introduced before, we obtain Eq.~\eqref{GWequi}. 

This result highlights a correspondence of the detected mode entanglement in two addressable modes and the detected particle entanglement in fully symmetric many-body quantum states. In the following we further generalize this criterion to an arbitrary number of modes $M$, and show how full multipartite inseparability can be detected with these methods.

\section{Generalization to multipartite entanglement}

When considering a collection of systems, entanglement can emerge in different partitions of the ensemble, \ie across any separation of the ensemble into groups of systems. Let us denote one specific partition as $\Lambda=\{\mathcal{A}_1,\dots,\mathcal{A}_k\}$, where the $\mathcal{A}$'s are non-overlapping groups of $1\leq|\mathcal{A}_q|\leq \Xi$ systems, such that $\sum_{q=1}^k|\mathcal{A}_q|=\Xi$. An $\Xi$-partite quantum state $\rho$ is called $\Lambda$-separable if it can be written as
\begin{align}\label{eq:Lsep}
    \rho_{\Lambda-\mathrm{sep}}=\sum_{\gamma}p_{\gamma}\rho_{\gamma}^{(\mathcal{A}_1)}\otimes\cdots\otimes\rho_{\gamma}^{(\mathcal{A}_{k} )} \;,
\end{align}
where the $\rho_{\gamma}^{(\mathcal{A}_q)}$ are quantum states of the subsystem $\mathcal{A}_q$. For an overview of different classes of entangled states in multipartite systems, we refer to Appendix~\ref{app:ent}.

\subsection{Inseparability of $M$ modes}
The entanglement criterion~(\ref{Giova}) can be generalized to yield a criterion for $\Lambda$-separable states of an $M$-mode system as follows~\cite{GessnerQuantum}: Any $\Lambda$-separable state must satisfy
\begin{equation}\label{GiovaM}
    \mathcal{G}_{\Lambda}^M(\vec{g},\vec{h})^2 := \dfrac{ \var\left[\sum_{I=1}^M g_I S_z^{I}\right] \var\left[\sum_{I=1}^M h_I S_y^{I}\right]}{\mathcal{B}_{\Lambda}^M(\vec{g},\vec{h})^2} \geq 1 \;,
\end{equation}
where
\begin{align}\label{Bk}
    \mathcal{B}_{\Lambda}^M(\vec{g},\vec{h}):=\frac{1}{2}\sum_{q=1}^l\left|\sum_{I\in\mathcal{A}_q}g_Ih_I\langle S^{I}_z\rangle\right| \;.
\end{align}
This bound holds for arbitrary choices of the coefficient vectors $\vec{g}=(g_1,\dots,g_M)$ and $\vec{h}=(h_1,\dots,h_M)$, which can be optimized to obtain the strongest possible witness. A violation of Eq.~\eqref{GiovaM} witnesses inseparability in the partition $\Lambda$.

We may further exclude separability in all partitions $\Lambda$ that contain at most $k$ subsystems by observing a violation of the single condition
\begin{align}\label{Gk}
    \mathcal{G}_{k}^M(\vec{g},\vec{h}):=\max_{\Lambda:\:l\geq k}\mathcal{G}_{\Lambda}^M(\vec{g},\vec{h}) \geq 1,
\end{align}
where the maximization runs over all partitions that consist of $l\geq k$ subsystems. A violation of the above bound with $k=M$, where each mode is treated as an individual subsystem, indicates that there is entanglement somewhere in the system without specifying how many subsystems are entangled. If the bound is violated for $k=2$, this means that we cannot identify even two separable groups, and we must thus consider the ensemble of all spins as a single entangled system. We remark here that this criterion analyzes each partition on a one-by-one basis, but it does not exclude arbitrary mixtures of separable models for different partitions, which is known as genuine multipartite entanglement~\cite{ReidTeh,HyllusEisert,GuehneToth} (see Appendix~\ref{app:ent} for details).

It is evident that the computation of the bound~(\ref{Gk}) becomes very demanding since the number of possible partitions increases exponentially with $M$. Moreover, identifying a suitable choice for the $\{\vec{g},\vec{h}\}$ introduces additional complexity.
A special case of Eq.~\eqref{GiovaM} is obtained for the choice of $\{\vec{g},\vec{h}\}$ given by
\begin{equation}\label{ghStar}
  g_I^{\ast}=1 \;,\qquad h_1^{\ast}=1 \;,\qquad h_{J}^{\ast}=-\frac{1}{M-1} \;,
\end{equation}
for all $I=1,\dots,M$ and $J=2,\dots,M$. 
With this choice we note that $g_1^{\ast} h_1^{\ast} = 1$, and $g_{I}^{\ast} h_{I}^{\ast} = -(M-1)^{-1}$ for $I>1$. 

Since for the symmetric spin states considered here, the variances in Eq.~(\ref{GiovaM}) do not depend on the partition $\Lambda$, the maximization in Eq.~(\ref{Gk}) affects only the bound~(\ref{Bk}). We thus obtain that $\mathcal{G}_{k}^M(\vec{g},\vec{h})=\mathcal{G}_{\Lambda_{\min}}^M(\vec{g},\vec{h})$, where $\Lambda_{\min}$ is the partition that achieves the minimum
\begin{align}\label{beta}
\beta^M_{k}(\vec{g},\vec{h})=\frac{|\langle S_z^A\rangle|}{2}\min_{\Lambda: l\geq k}\beta^M_{\Lambda}(\vec{g},\vec{h}) \;,
\end{align}
and $\beta^M_{\Lambda}(\vec{g},\vec{h}):=\sum_{q=1}^l\left|\sum_{I\in\mathcal{A}_q}g_Ih_I\right|$. In writing Eq.~(\ref{beta}), we made use of the symmetry property~(\ref{evS}) to limit the optimization procedure to the coefficients $\{\vec{g},\vec{h}\}$. Next, we observe that all contributions of terms from sets with $I>1$ will increase $\beta^M_{\Lambda}(\vec{g}^*,\vec{h}^*)$ whenever they appear in a partition $\Lambda$ that distinguishes them from mode $I=1$, whereas these terms will decrease $\beta^M_{\Lambda}(\vec{g}^*,\vec{h}^*)$ when in a partition $\Lambda$ that lumps them into the set $\mathcal{A}_1$ together with mode $1$. From this argument we also see that it is advantageous to pick a partition that splits the system in as few subsystems as possible. Since for a given $k$, at least $k$ subsystems must be formed, the optimal partition describes $k-1$ single-mode subsystems (with $I>1$) and places all other modes (including $I=1$) into a single subsystem. The minimum bound is thus given by
\begin{align}
\beta^M_{\Lambda}(\vec{g}^*,\vec{h}^*)&=\vert g_1^{\ast} h_1^{\ast} \underbrace{ + \dots+g_{M-(k-1)}^{\ast} h_{M-(k-1)}^{\ast}}_{M-k \;\text{terms}} \vert\notag\\&\quad+\underbrace{ \vert g_{M-(k-2)}^{\ast} h_{M-(k-2)}^{\ast} \vert + \cdots + \vert g_M^{\ast} h_M^{\ast} \vert }_{k-1 \;\text{terms}}\notag\\
& =\dfrac{2(k-1)}{M-1}\;.
\end{align}
In summary, the minimum bound at the denominator of Eq.~\eqref{GiovaM} takes the form
\begin{equation}\label{eq:BkM}
    \mathcal{B}_k^M(\vec{g}^\ast,\vec{h}^\ast) = \dfrac{1}{2}\left( \dfrac{2(k-1)}{M-1} \right)\abs{\avg{S_x^A}} \;.
\end{equation}

The choice given in Eq.~\eqref{ghStar} gives for the variances
\begin{equation}\label{eq:Vargisi}
    \var\left[\sum_{I=1}^M g_I S_z^{I}\right] = \var\left[ S_z \right] \;,
\end{equation}
and, using again Eqs.~(\ref{evS}) and (\ref{evSS}) with $\avg{\PX}=1/M$, that
\begin{equation}\label{varY}
    \var\left[\sum_{I=1}^M h_I S_y^{I}\right] = \dfrac{N}{4(M-1)} \;.
\end{equation}
A detailed calculation is given in Appendix~\ref{app:VyG}.

\begin{figure}
  \centering
\includegraphics[width=0.46\textwidth]{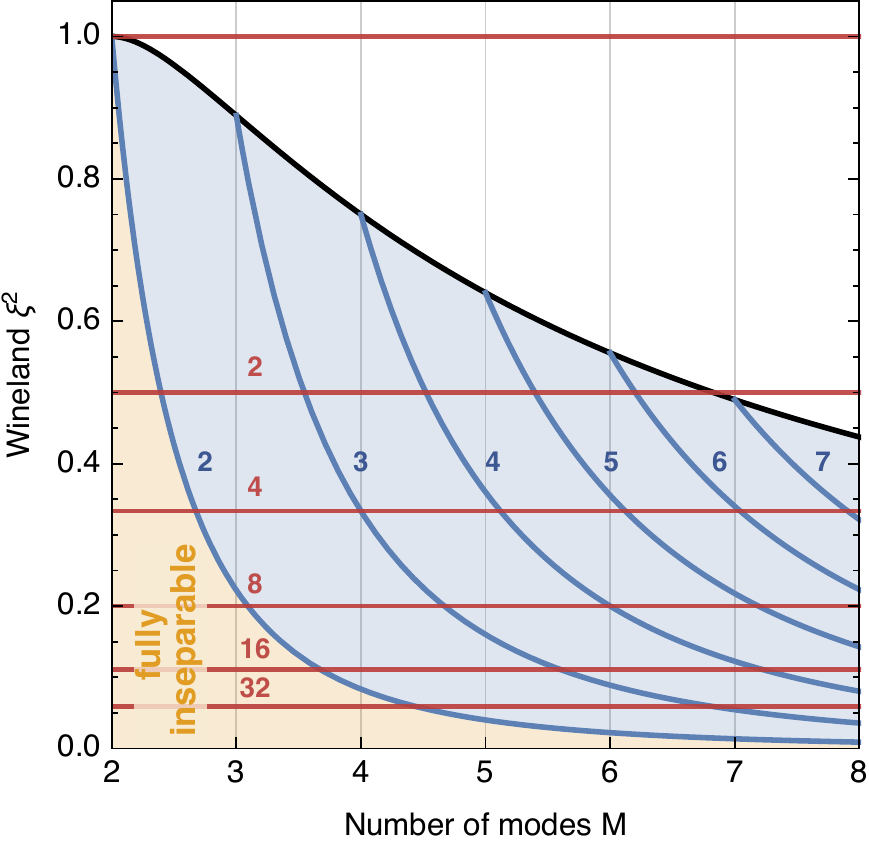}
  \caption{Bounds for mode and particle entanglement from collective spin measurements. We compare the Wineland \textit{et al.} spin-squeezing coefficient to the limits on particle entanglement~\eqref{WkprodTight} (red lines) and mode entanglement for splitting into $M$ modes~\eqref{kSepGCrit} (blue lines) as a function of $M$. For values of $\xi^2$ below the top black line (where $k=M$), mode entanglement is revealed. Upon crossing additional blue lines, the entanglement does not allow us to split the system into more than $k$ separable sets of modes, where $k$ is indicated in blue next to the lines. The yellow region corresponds to $k=2$, and it indicates where all modes must be treated as one entangled entity that cannot be partitioned into separable groups.}
    \label{fig:Mkp}
\end{figure}

Using the definition of $\xi^2$~(\ref{Wine}), together with Eqs.~(\ref{eq:BkM}), (\ref{eq:Vargisi}) and (\ref{varY}), we can express Eq.~(\ref{GiovaM}) as
\begin{align}
    \mathcal{G}_k^M(\vec{g}^\ast,\vec{h}^\ast)^2 
    &= \xi^2 \dfrac{M^2(M-1)}{4 (k-1)^2} \geq 1 \;. \label{GWequiM}
\end{align}
From this, we conclude that any state that is separable into $k$ subsystems or more must satisfy
\begin{align}
\xi^2 \geq \dfrac{4 (k-1)^2}{M^2(M-1)} \;. \label{kSepGCrit}
\end{align}
Therefore, observing $\xi^2 < 4 (k-1)^2/(M^2(M-1))$ implies more than $k$ partite inseparability (see blue lines in Fig.~\ref{fig:Mkp}). This is the main result of this section. It implies, \eg that mode entanglement $(k=M)$ is observed among $M$ modes whenever $M < 2( 1 + \sqrt{1-\xi^2})/\xi^2$ (black line in Fig.~\ref{fig:Mkp}).

Since any state can be considered as a single indivisible system, the bound becomes trivial for $k=1$ and it can never be violated in this case. Generally, meaningful values for $k$ range from $2$ to $M$, and the smaller $k$ is, the more modes are recognized as entangled. If the bound is violated for $k=2$ this implies that there is no separable partition at all, and hence all $M$ modes must be entangled. For $M=2$, the criterion Eq.~\eqref{GWequiM} reduces to Eq.~\eqref{GWequi}, as expected. 

We recall that our conclusions are based on specific entanglement witnesses, \ie sufficient conditions for entanglement. Hence, these results only put a lower bound on the actual number of entangled modes.

\subsection{Limits on global and local spin-squeezing}
Let us now investigate the lower bound for the spin-squeezing coefficient $\xi^2$. As we show in Appendix~\ref{app:xilimit}, an arbitrary spin-$S$ system always satisfies the bound
\begin{equation}\label{eq:minxi2S}
    \xi^2\geq \frac{1}{1+S} \;,
\end{equation}
where the equality can be approached asymptotically.

Furthermore, we can define the local spin-squeezing coefficient
\begin{equation}\label{WineLoc}
\xi^2_{I} := \dfrac{N^I \var\left[S_z^I\right]}{\vert\langle S_x^I\rangle\vert^2} \;,
\end{equation}
and show that there exists a limit on the squeezing that can be achieved locally from the splitting of a symmetric squeezed state. Under the assumptions (i), (ii) and (iiib) of Sec.~\ref{sec:symmetry}, the local squeezing obeys the bound 
\begin{equation}\label{locB}
    \xi^2_{I} \geq 1 - \avg{\PX} \;,
\end{equation}
where the equality can be approached asymptotically in the limit $S\rightarrow\infty$ (see Appendix~\ref{app:LCLxilimit}). Let us emphasize that in the derivation of Eq.~\eqref{locB} we did not use assumption (iiia), meaning that the inequality holds even for asymmetric splittings into $M$ modes, \ie for more general cases where $\avg{\PX}$ depends on $I$.

To conclude, we can also show that there is an exact relation between the global and the local spin-squeezing coefficients, namely (see Appendix~\ref{app:GLxiRelation})
\begin{equation}\label{eq:GlobLoc}
\xi^2 = \sum_{I=1}^M \xi^2_I  - \dfrac{N^2 (M-1)}{4 \avg{S_x}^2} \;.
\end{equation}
Also here, analogously to Eq.~\eqref{locB}, it is worth emphasizing that Eq.~\eqref{eq:GlobLoc} holds even for asymmetric splitting where $\avg{\PX}$ depends on $I$. However, in the case where $\avg{\PX}=1/M$, Eq.~\eqref{eq:GlobLoc} can be used in conjunction with Eq.~\eqref{kSepGCrit} to relate local squeezing and collective polarization to mode inseparability.

\subsection{Multipartite entanglement detection from spin squeezing}

\subsubsection{State-independent multipartite entanglement bounds}
Spin squeezing provides quantitative bounds on the number of entangled particles from collective measurements. Furthermore, the detected entanglement is relevant for the improvement of measurement precision in quantum metrology. To see this, recall that the quantum Fisher information $\mathcal{F}_Q[\rho,H]$~\cite{BraunsteinCaves} quantifies the metrological sensitivity of a quantum state $\rho$ under an evolution generated by the Hamiltonian $H$~\cite{HelstromBook,GiovannettiReview,RMP}. By virtue of the inequality~\cite{PS09}
\begin{align}\label{eq:xiFQ}
    \xi^{-2}\leq \frac{\mathcal{F}_Q[\rho,S_y]}{N},
\end{align}
the inverse spin-squeezing coefficient $\xi^{-2}$ can be interpreted as a Gaussian approximation to the full sensitivity, normalized by the total number of particles~\cite{GessnerPRL2019}. The detection of metrologically useful entanglement makes use of the fact that $p$-producible $N$-qubit quantum states (\ie states that contain at most $p$ entangled particles; see Appendix~\ref{app:ent}) can only achieve sensitivites up to $\mathcal{F}_Q[\rho_{p},S_y]\leq pN$~\cite{PS09,HyllusToth}. Combining this bound with the inequality~(\ref{eq:xiFQ}), we find that a violation of~\cite{GessnerPRA2017}
\begin{align}\label{Wkprod}
\xi^2\geq \frac{1}{p} \;,
\end{align}
implies the presence of entanglement among more than $p$ particles (see blue lines in Fig.~\ref{fig:MSF}). For non-integer $N/p$ a small improvement of this bound can be achieved using a more general expression~\cite{HyllusToth}.

Interestingly, we can derive a much tighter bound than Eq.~\eqref{Wkprod}. This is possible because the limit $1/p$ arises from the bound on the quantum Fisher information that is achieved only by products of maximally entangled Greenberger-Horne-Zeilinger (GHZ) states~\cite{HyllusToth}, for which $\xi^2$ actually diverges, so that Eq.~\eqref{Wkprod} can never be saturated. Instead, by making use of the asymptotically achievable limit~(\ref{eq:minxi2S}), we can show that a violation of
\begin{equation}\label{WkprodTight}
    \xi^2\geq \frac{1}{1+p/2}
\end{equation}
implies entanglement of more than $p$ spins among the total number of $N$ spins-$1/2$ particles. This is the central result of this section and it follows as a consequence of convexity and subadditivity properties of the inverse spin-squeezing coefficient $\xi^{-2}$. The details can be found in Appendix~\ref{app:tighterbound}, where we also demonstrate that for non-integer $N/p$, the bound~\eqref{WkprodTight} can be improved to the expression:
\begin{equation}\label{WkprodTightNp}
    \xi^{2}\geq \frac{N}{N_p \frac{p^2}{2} +\frac{r^2}{2}+N},
\end{equation}
with $N_p=\lfloor N/p\rfloor$ and $r=N-pN_p$.
We emphasize that, contrary to Eq.~\eqref{Wkprod}, this bound can be (asymptotically) saturated, and for $p$ large it is higher than Eq.~\eqref{Wkprod} by a factor of two (see red solid lines in Fig.~\ref{fig:MSF}). 

In a system with $p$ spin-$1/2$ particles, the bound~(\ref{WkprodTight}) can be approached asymptotically in the limit of infinite squeezing and vanishing polarization $\avg{S_x}$. Such states are known as Twin-Fock states $|\Psi_{\rm TF}\rangle$ (see, e.g., Ref.~\cite{Klempt} for an experimental study of their metrological entanglement) and the bound~(\ref{WkprodTight}) expresses their full sensitivity as quantified by the quantum Fisher information $\mathcal{F}_Q[|\Psi_{\rm TF}\rangle,S_y]=p(1+p/2)$ [compare Eqs.~(\ref{WkprodTight}) and~(\ref{eq:xiFQ})]. By quantifying the maximum sensitivity achievable with Gaussian measurements, the result~(\ref{WkprodTight}) implies that any sensitivity of $p$ spin-$1/2$ particles that exceeds this bound, \ie any state with $\mathcal{F}_Q[\rho,S_y]>p(1+p/2)$, must necessarily be non-Gaussian in the sense that its metrological features cannot be captured through spin-squeezing coefficients.

\subsubsection{State-dependent multipartite entanglement bounds}
\begin{figure}
  \centering
\includegraphics[width=0.46\textwidth]{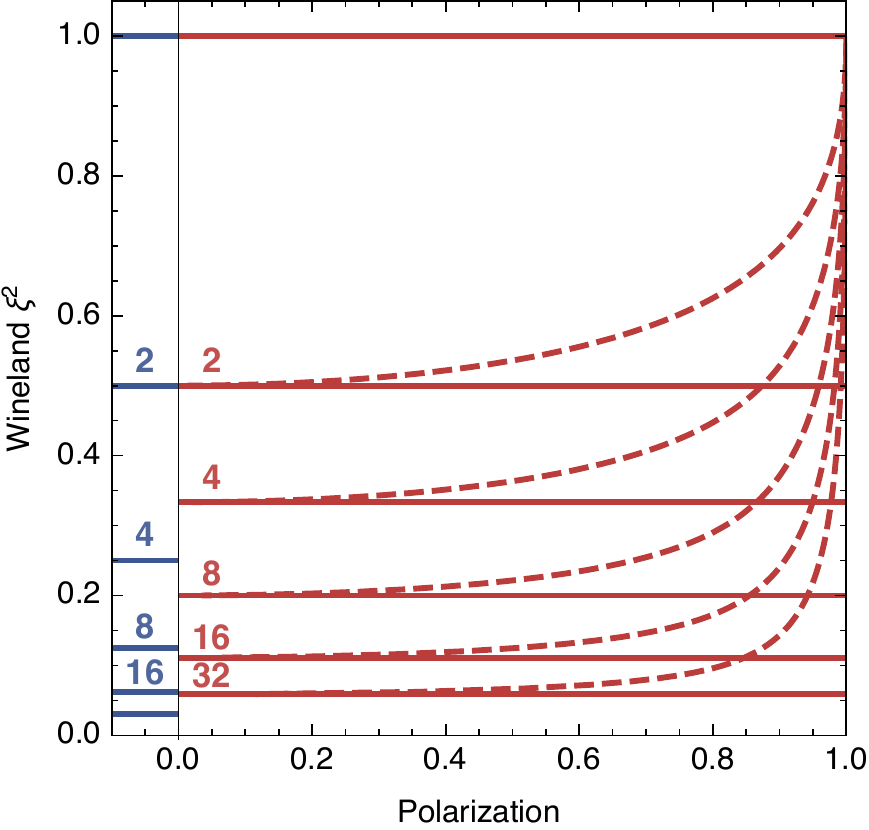}
  \caption{Detecting multiparticle entanglement from spin squeezing. A number $p$ of entangled particles is detected when the Wineland \textit{et al.} spin-squeezing coefficient is lower than the respective lines. The solid blue lines are the constant bounds obtained from the Fisher information, Eq.~\eqref{Wkprod}. The solid red lines are improved bounds obtained from minimizing $\xi^2$ for fixed $S=p/2$, Eq.~\eqref{WkprodTight}. These bounds are independent of the polarization $\avg{S_x}/S$. In contrast, the dashed lines are state-dependent bounds obtained from the S{\o}rensen-M{\o}lmer relation Eq.~\eqref{MS}, where $\xi^2$ is minimized numerically for a fixed $S=p/2$ and polarization \cite{SorensenMolmer}. The state-independent bounds~\eqref{WkprodTight} can be recovered from this approach in the limit of vanishing contrast.}
    \label{fig:MSF}
\end{figure}

In practice, in order to access $\xi^2$ one actually measures $\var[S_z]$ and $\avg{S_x}$ separately, rather than the ratio $\var[S_z]/\avg{S_x}^2$ itself. Having independent knowledge of these two quantities, it is possible to construct a stronger multipartite entanglement witness than Eq.~\eqref{WkprodTight}. S\o{}rensen and M\o{}lmer~\cite{SorensenMolmer} showed that states with no more than $p$-partite entanglement satisfy
\begin{align}\label{MS}
\var\left[ S_z \right]&\geq
     S \, F_{S_p}\left[\frac{\langle S_x\rangle}{S}\right],
\end{align}
where $S_p=p/2$, and the functions $F_{S}[x]$ are obtained (\eg numerically) by minimizing the variance $\var[S_z]$ of a spin $S$ as a function of its mean spin $\avg{S_x}$~\cite{SorensenMolmer}. This approach is constructed such that it detects the largest family of entangled states on the basis of $\var[S_z]$ and $\avg{S_x}$. However, since the metrological sensitivity is determined only by the ratio of these two quantities, the multipartite entanglement detected by this approach is not immediately linked to a metrological advantage. Yet, the criterion~(\ref{MS}) is more powerful than Eq.~(\ref{WkprodTight}), since it makes use of the additional information provided by $\avg{S_x}$. Indeed, we demonstrate in Appendix~\ref{app:MSWINE} that in the limit $\avg{S_x}\to 0$, the condition~(\ref{MS}) coincides with~(\ref{WkprodTight}). Since this corresponds to the limit in which the criterion~(\ref{MS}) is least effective, we can interpret this limit as ignoring the additional information that is provided by the mean spin length, assuming the worst-case scenario. This can be seen in Fig.~\ref{fig:MSF}, where we compare the constant bound for $p$-partite entanglement obtained from Eq.~\eqref{WkprodTight} (red continuous lines) to the state-dependent bound from Eq.~\eqref{MS} (red dashed lines).

We show in Appendix~\ref{app:SM} how condition~\eqref{MS} can be improved for non-integer $N/p$, and how the resulting expression reproduces the bound~(\ref{WkprodTightNp}) in the limit of vanishing polarization $\avg{S_x}$. Condition~\eqref{MS} also allows to identify genuine $p$-partite entanglement \cite{ReidHeDrummundFrontiers2011}, meaning that one can exclude convex combinations of $(p-1)$-producible states (see Appendix \ref{app:ent} and \ref{app:genpMS}). Moreover, it can be generalized to systems with fluctuating particle numbers~\cite{Hyllus2}.

\section{Relation with two-way EPR steering}
For a bipartite scenario ($M=2$) the criterion Eq.~\eqref{Giova} can be extended to detect also a stronger form of entanglement, namely EPR steering. Specifically, states that do not allow for steering of system $B$ by $A$ satisfy the condition~\cite{CavalcantiReid,Bowen,ReidRMP}
\begin{equation}\label{GiovaS}
\mathcal{R}^2 := \dfrac{4 \var\left[S_z^A + S_z^B \right] \var\left[S_y^A - S_y^B \right] }{\vert\langle S_x^B\rangle\vert^2} \geq 1 \;.
\end{equation}
Therefore, a violation of Eq.~\eqref{GiovaS} reveals steering of $B$ by $A$. A similar criterion holds for steering of $A$ by $B$.

In the following we will focus on the symmetric scenario, where measurements in system $A$ and $B$ yield the same results. In this case we have $\avg{ S_x^A}=\avg{S_x^B}$ which allows us to express the condition~(\ref{GiovaS}) equivalently as
\begin{equation}\label{RGW}
\mathcal{R}^2 = 4\mathcal{G}^2 = 4\xi^2 \geq 1 \;.
\end{equation}
Because of symmetry, a violation of this relation directly implies two-way steering between modes $A$ and $B$.

Combined with our results from the previous section, we conclude that if we want to observe steering through Eq.~\eqref{GiovaS}, we need to satisfy the condition
$\xi^2<1/4$, which implies entanglement of $p>6$ particles. However, note that Eq.~\eqref{GiovaS} can be generalized by including free coefficients in front of the spin operators, similarly to Eq.~\eqref{GiovaM}. This allows to detect EPR correlations with less squeezing \cite{FadelSplitBEC}, but the correspondence given in Eq.~\eqref{RGW} is lost.

Interestingly, a bipartite EPR criterion can also be derived from the S\o{}rensen-M\o{}lmer bounds Eq.~\eqref{MS}. We show in Appendix~\ref{app:SMepr} that a violation of
\begin{equation}\label{eq:eprMS}
    \var[S_z] \geq S_B \, F_{S_B}\left[ \dfrac{\langle S_x^{B} \rangle }{S_B} \right]
\end{equation}
implies steering of $B$ by $A$. This criterion is easier to violate than the condition $\var[S_z] \geq S \, F_{S_B}\left[ \dfrac{\langle S_x^{B} \rangle }{S} \right]$ that was derived in Ref.~\cite{ReidHeDrummundFrontiers2011}. However, it is still very demanding to witness steering with this approach, since no assumptions can be made about the properties of system $A$.

\vspace{5mm}

\section{Discussions and conclusions}
In this work we established relations between criteria for multipartite entanglement of particles and modes based on the measurement of first and second moments of collective spin observables. In the case of symmetric spin states, we found that the Wineland \textit{et al.} spin-squeezing coefficient~\cite{Wineland} coincides with a witness of mode entanglement that is based on Heisenberg-Robertson-type uncertainty relations with modified bounds~\cite{Giovannetti}. This correspondence can be extended to reveal a direct relation between the spin-squeezing coefficient of symmetric spin states and a two-way EPR-steering criterion of two addressable modes.

We further revealed the relation between different multipartite entanglement criteria based on spin squeezing. The Wineland \textit{et al.} spin-squeezing coefficient~\cite{Wineland} captures the metrological sensitivity gain and can be used to study multiparticle entanglement. The approach by S\o{}rensen and M\o{}lmer~\cite{SorensenMolmer} makes use of the independent knowledge of the spin polarization to derive optimized state-dependent bounds on the spin-squeezing coefficient for multipartite entangled states. Alternatively, state-independent bounds can be derived by exploiting the relation between spin squeezing and the Fisher information~\cite{PS09}, but these bounds are not saturable by Gaussian states~\cite{HyllusToth}. We addressed this limitation by deriving state-independent bounds that can be asymptotically saturated. This provides the tightest state-independent bounds on the spin-squeezing coefficient for the detection of multipartite entangled states. Interestingly, we observe that these bounds coincide with those of S\o{}rensen and M\o{}lmer \cite{SorensenMolmer} in the limit of vanishing polarization.

Moreover, we identified a simple expression for the maximum spin squeezing that can be achieved locally from the splitting of a squeezed state. Our results provide bounds on the amount of addressable multimode entanglement that can be generated by distributing identical particles into external modes. For example, they apply to nonclassical states of BECs that are split into different spatial modes, as in Refs.~\cite{FadelSplitBEC,KunkelSplitBEC,LangeSplitBEC}.

\section{Acknowledgments}
We thank Qiongyi He, Margaret Reid, Run Yan Teh and Philipp Treutlein for useful discussions. MF acknowledges support by the Swiss National Science Foundation. MG acknowledges funding by the LabEx ENS-ICFP: ANR-10-LABX-0010/ANR-10-IDEX-0001-02 PSL*.

\clearpage

\begin{widetext}

\appendix

\section*{Supplementary material}

Here we show the detailed calculations for the results presented in the paper.

\section{Detailed calculations for proving Eq.~\eqref{GWequi}}

\subsection{Proof of Eq.~\eqref{evSS}}\label{app:A1corr}

\begin{subequations}
\begin{align}
\avg{ \SuX \SvY } &= \sum_{i,j=1}^N \avg{\sui \svj \PXi \PYj}  \\
&= \sum_{i=1}^N \avg{\sui \svi \PXi \PYi} + \sum_{\substack{i,j=1\\i\neq j}}^N \avg{\sui \svj \PXi \PYj}  \\
&= \sum_{i=1}^N \avg{\sui \svi} \avg{\PXi \PYi} + \sum_{\substack{i,j=1\\i\neq j}}^N \avg{\sui \svj} \avg{\PXi \PYj}  &\;\text{using (i)} \\
&= \sum_{i=1}^N \avg{\sui \svi} \delta_{I,J} \avg{\PXi} + \sum_{\substack{i,j=1\\i\neq j}}^N \avg{\sui \svj} \avg{\PXi \PYj}  &\;\text{projectors are orthogonal} \\
&= \delta_{I,J} \avg{\PX} \sum_{i=1}^N \avg{\sui \svi} + \avg{\PX} \avg{\PY}  \sum_{\substack{i,j=1\\i\neq j}}^N \avg{\sui \svj}  &\;\text{using (ii, iiib)} \\
&= \delta_{I,J} \avg{\PX} N \avg{\su^{(1)} \sv^{(1)}} + \avg{\PX} \avg{\PY} N(N-1) \avg{\su^{(1)} \sv^{(2)}}  \;.
\end{align}
\end{subequations}

Moreover, if $\vec{u}=\vec{v}$ we obtain
\begin{equation}
\avg{ \SuX \SuY } = \delta_{I,J} \avg{\PX} N \dfrac{1}{4} + \avg{\PX} \avg{\PY} N(N-1) \avg{\su^{(1)} \su^{(2)}} 	\;.
\end{equation}

\subsection{Proof of Eq.~\eqref{VyG}}\label{app:A2VarY}
Let us first express the variance and covariance of collective spins under the assumptions of symmetry. We obtain
\begin{align}
\var\left[ \SuA \right] &= \avg{(\SuA)^2} - \avg{\SuA}^2 \nonumber\\
&= \left( \avg{\PA} N \dfrac{1}{4} + \avg{\PA}^2 N(N-1) \avg{\su^{(1)} \sv^{(2)}}  \right) - \left( \avg{\PA} N \avg{\su^{(1)}} \right)^2 \label{eq:VarSA}
\end{align}
and
\begin{align}
\cov\left[ \SuA, \SuB \right] &= \avg{\SuA \SuB} - \avg{\SuA}\avg{\SuB} \nonumber\\
&= \left( \avg{\PA}\avg{\PB} N(N-1) \avg{\su^{(1)} \sv^{(2)}}  \right) - \left( \avg{\PA} N \avg{\su^{(1)}} \right) \left( \avg{\PB} N \avg{\su^{(1)}} \right) \nonumber\\
&= \left( \avg{\PA}^2 N(N-1) \avg{\su^{(1)} \sv^{(2)}}  \right) - \left( \avg{\PA} N \avg{\su^{(1)}} \right)^2.
\end{align}
Combining these expressions, we obtain
\begin{align}
    \var\left[S_y^A - S_y^B \right] &= \var\left[S_y^A \right] + \var\left[S_y^B \right] - 2 \cov\left[S_y^A, S_y^B \right] \nonumber\\
    &= 2 \langle\PA\rangle \dfrac{N}{4} \;.
\end{align}

\section{Definitions of multipartite entanglement}\label{app:ent}
We briefly review different inequivalent notions of entanglement in multipartite systems. The definition provided in Eq.~\eqref{eq:Lsep} of the main text describes separability in a specific partition $\Lambda=\{\mathcal{A}_1,\dots,\mathcal{A}_l\}$, where each of the $\mathcal{A}_q$ is a group of $|\mathcal{A}_q|$ systems. Such a partition $\Lambda$ can be characterized either by the size of its largest group $w(\Lambda):=\max_q|\mathcal{A}_q|$ or by the number of groups it contains $h(\Lambda):=l$ (these two quantities correspond to width $w$ and height $h$ of the Young diagram associated with $\Lambda$~\cite{SzalayQuantum}). Separable models with $w(\Lambda)\leq p$ are called $p$-producible and those with $h(\Lambda)\geq k$ are called $k$-separable. These definitions can be applied to classify the type of correlations in the context of particle entanglement, where each particle is considered as a system, and mode entanglement, where each mode is considered as a system.

In the context of this paper, we provide criteria for ``$k$-inseparable'' states of modes, \ie states that are incompatible with all mode-separable models~(\ref{eq:Lsep}) with $h(\Lambda)\geq k$. In the above definition, all separable models are excluded individually, \ie we verify incompatibility with all descriptions of the kind~(\ref{eq:Lsep}) for each $\Lambda$ in $\mathfrak{L}_{k-\mathrm{sep}}=\{\Lambda\:|\: h(\Lambda)\geq k\}$. A stronger condition requires the exclusion of all convex combinations of a specific family of separable models, and it is usually emphasized by the term ``genuine'', see \eg Ref.~\cite{ReidTeh}. For instance, we would call a state genuine $k$-inseparable if it is incompatible with any description of the kind
\begin{align}\label{eq:genuineENT}
    \rho=\sum_{\Lambda\in\mathfrak{L}_{k-\mathrm{sep}}}P_{\Lambda}\rho_{\Lambda-\rm{sep}},
\end{align}
where $P_{\Lambda}$ is a probability distribution and the $\rho_{\Lambda-\rm{sep}}$ are of the form~(\ref{eq:Lsep}). In the context of particle entanglement, we further say that a state has [genuine] $p$-partite entanglement if it excludes all [convex combinations of] $(p-1)$-producible models (recall that $p$-producible models are described by the set $\mathfrak{L}_{p-\mathrm{prod}}=\{\Lambda\:|\: w(\Lambda)\leq p\}$).

\section{Detailed calculations for proving Eq.~\eqref{GWequiM}}\label{app:VyG}

\subsection{Proof of Eq.~\eqref{beta}}
For a given $k$, one has to find the minimum among the bounds given by each partition. However, we can easily see that, because of the signs, the minimum bound will come from terms of the form
\begin{equation}
\mathcal{B}(M,k) = \vert g_1^{\ast} h_1^{\ast} \underbrace{ - g_2^{\ast} h_2^{\ast} - \cdots}_{M-k \;\text{terms}} \vert + \underbrace{ \vert g_I^{\ast} h_I^{\ast} \vert + \cdots + \vert g_M^{\ast} h_M^{\ast} \vert }_{k-1 \;\text{terms}} \;.
\end{equation}
This is because $g_1^{\ast} h_1^{\ast} = 1$ is the largest term, that can be minimized by subtracting as many terms $g_{I}^{\ast} h_{I}^{\ast}$ with $I>0$ as possible.

Our choices give, for $0\geq c,d \geq M-1$
\begin{subequations}
\begin{align}
\vert g_1^{\ast} h_1^{\ast} + c\;  g_2^{\ast} h_2^{\ast} 	\vert &=  \dfrac{M-1-c}{M-1} \\
\vert d\;  g_2^{\ast} h_2^{\ast} \vert &=  \dfrac{d}{M-1}
\end{align}
\end{subequations}
and, with $c=M-k$, $d=k-1$, we get $\beta_k^M$.

\subsection{Proof of Eq.~\eqref{varY}}

\begin{align}
\var\left[ \sum_{I=1}^M h_I \SuX \right] 
&= \sum_{I,J=1}^M \cov\left[ h_I \SuX, h_J \SuY \right]  \nonumber\\
&= \sum_{I=1}^M h_I^2 \var\left[ \SuX \right] + \sum_{I\neq J=1}^M h_I h_J \cov\left[ \SuX, \SuY \right]  \nonumber\\
&= T_1 \left(\sum_{I=1}^M h_A^2\right) + T_2 \left( \sum_{I=1}^M h_I^2 + \sum_{I\neq J=1}^M h_I h_J \right)
\end{align}
with
\begin{align}
    T_1 &= \avg{\PX} \dfrac{N}{4} \\
    T_2 &= \avg{\PX}^2 N(N-1) \avg{\su^{(1)} \sv^{(2)}}  -  \left( \avg{\PX} N \avg{\su^{(1)}} \right)^2 \;.
\end{align}

With the choice $h_1^{\ast} = 1$ and $h_{I}^{\ast} = (M-1)^{-1}$ for $I>1$, we have
\begin{equation}
\sum_{I=1}^M {h_I^{\ast}}^2 = \dfrac{M}{M-1}  \;,\qquad   \sum_{I\neq J=1}^M h_I^{\ast} h_J^{\ast} = - \dfrac{M}{M-1}
\end{equation}
and, therefore, taking $\avg{\PX}=1/M$, we have
\begin{equation}
\var\left[ \sum_{I=1}^M h_I^{\ast} \SuX \right] = \dfrac{N}{4(M-1)} \;,
\end{equation}
a result that is independent of the state and of the direction $\vec{u}$.

\section{Limits on spin squeezing}
\subsection{Ultimate limit on spin squeezing}\label{app:xilimit}
For a given, integer spin length $S$, we determine the minimum possible value of $\xi^2$ attainable by any quantum state. Our aim is to identify states with minimal $\var{[S_z]}$ with fixed $\avg{S_x}$. S\o{}rensen and M\o{}lmer~\cite{SorensenMolmer} pointed out that these states satisfy $\avg{S_z}=0$ and can therefore be found as the ground states of the Hamiltonian $H=\lambda S_x + S_z^2$, where $\lambda$ takes on the role of a Lagrange multiplier, see also~\cite{RojoPRA2003}. We use first-order perturbation theory to determine the value of $\xi^2$ for the ground state in the presence of some small but finite $\lambda$, and then take the limit $\lambda\to 0$. We start from the ground state of the (unperturbed) Hamiltonian $H_0=S_z^2$, which is
the state $\ket{S,m_z}=\ket{S,0}$. Considering $S_x$ as the perturbation, the ground state of $H$ to first order in $\lambda$ is (up to normalization)
\begin{align}
    \ket{\tilde{\psi}} &= \ket{S,0} + \lambda \sum_{m\neq 0} \dfrac{\bra{m,S}S_x\ket{S,0}}{E_0 - E_m} \ket{S,m} \\
    &= \ket{S,0} - \lambda \dfrac{1}{2}\sqrt{S(S+1)}\left(\ket{S,-1} + \ket{S,+1} \right) \;,
\end{align}
where $E_m=\bra{S,m}H_0\ket{S,m}$, and given the form of $S_x$ the summation effectively runs only over $m=\pm 1$. We use this expression (plus normalization) to evaluate the expectation values
\begin{align}
    \avg{S_x} &= -\dfrac{2S(S+1)\lambda}{2+S(S+1)\lambda^2} \label{eq:pertSx}\\
    \avg{S_z^2} &= 1 - \dfrac{2}{2+S(S+1)\lambda^2} \;. \label{eq:pertSz2}
\end{align}
The spin-squeezing coefficient obtained from these expressions reads
\begin{equation}
    \xi^2 = \dfrac{1}{S+1} + \dfrac{S}{2}\lambda^2 \;.
\end{equation}
We conclude that, in any spin-$S$ system, we have
\begin{align}\label{eq:minxi}
\xi^2 \geq \dfrac{1}{S+1}.
\end{align}
As explained in the main text, this limit corresponds to the sensitivity of a twin-Fock state $\ket{S,0}$ (as expressed by its quantum Fisher information).

\subsection{Limit on local spin squeezing after splitting}\label{app:LCLxilimit}
Let us now identify the limit of the local spin-squeezing coefficient in one of the modes after splitting an ensemble described by the global coefficient $\xi^2$ into $M$ modes.
We can define the local spin-squeezing coefficient as
\begin{equation}
\xi_I^2 := \dfrac{N^I \var\left[S_z^I\right]}{\vert\langle S_x^I\rangle\vert^2}\;,
\end{equation}
where $N^I=N\avg{\PX}$ and, using Eqs.~(\ref{evS}) and~(\ref{eq:VarSA}), we find that the relation with the total squeezing $\xi^2$ is
\begin{equation}\label{eq:xiaExpr}
\xi_I^2 = \xi^2 \avg{\PX} + \left(1-\avg{\PX}\right)  \left(\dfrac{N\avg{\PX}/2}{\avg{ S_x^I}}\right)^2 \;.
\end{equation}

Because $\xi^2\geq 0$ and $N\avg{\PX}/2\avg{S_x^I} \geq 1$, we obtain that
\begin{equation}\label{eq:xiAlimit}
    \xi_I^2 \geq 1-\avg{\PX} \;.
\end{equation}
This expression tells us that, even if the intitial state comes close to the limit $\xi^2\rightarrow 0$ and $\avg{S_x^I}\rightarrow N\avg{\PX}/2$ (which can be approached in the limit $N\rightarrow \infty$ with optimized squeezing), after the splitting the squeezing will always be limited by~\eqref{eq:xiAlimit}. If $\avg{\PX}=1/2$, one obtains locally at most $-3\;\text{dB}$ of spin squeezing. If $\avg{\PX}=1/3$, $\xi_I^2 \geq -1.76\;\text{dB}$.

\subsection{Relation between global and local squeezing}\label{app:GLxiRelation}
To relate the global squeezing $\xi^2$ to the sum of local squeezing coefficients $\xi^2_I$, we make use of Eq.~\eqref{evS} to rewrite Eq.~\eqref{eq:xiaExpr} as:
\begin{equation}
\xi^2 \avg{\PX} = \xi_I^2 - \left(1-\avg{\PX}\right)  \left(\dfrac{N}{2\avg{ S_x}}\right)^2 \;.
\end{equation}
Summing both sides over all modes $I=1,\dots,M$, and using the fact that $\sum_{I=1}^M \avg{\PX}=1$, we obtain Eq.~(\ref{eq:GlobLoc}).

\section{Derivation of sharper multipartite entanglement bounds on $\xi^2$}\label{app:tighterbound}
Our derivation of the criterion for $p$-partite entanglement~(\ref{WkprodTight}) makes use of the ultimate limit on the spin-squeezing coefficient $\xi^2$ in arbitrary quantum states of a spin-$S$ system that was derived in Appendix~\ref{app:xilimit}. We further use  convexity and subadditivity of the inverse spin-squeezing coefficient, which we demonstrate below. Finally, we combine these results to derive the bound~(\ref{WkprodTight}) on $p$-producible states.

\subsection{Convexity and subadditivity of $(2S)\xi^{-2}$}
Consider an arbitrary linear combination of quantum states $\rho = \sum_{\gamma}p_{\gamma}\rho_{\gamma}$. The inverse spin-squeezing coefficient satisfies the convexity property~\cite{GessnerPRL2019}
\begin{align}\label{eq:convexity}
    \xi^{-2}_{\rho}\leq \sum_{\gamma}p_{\gamma}\xi^{-2}_{\rho_{\gamma}}.
\end{align}
Let us demonstrate that $(2S)\xi^{-2}_{\rho}$ is also subadditive, where $S$ is the total spin of the system described by $\rho$. To this end, we consider a product state $\rho=\rho_1\otimes\cdots\otimes\rho_M$ and we decompose the total spin $S$ into its local components as $\Su=\sum_{I=1}^M \SuX$. We write
\begin{align}
    (2S)\xi^{-2}_{\rho}=\frac{\avg{S_x}_{\rho}}{\var{[S_z]}_{\rho}},
\end{align}
and the absence of correlations in $\rho$ implies that
\begin{subequations}
\begin{align}
    \avg{S_x}_{\rho}&=\sum_{I=1}^M\avg{S_x^{I}}_{\rho_I}\\
    \var{[S_z]}_{\rho}&=\sum_{I=1}^M\var{[S_x^{I}]}_{\rho_I}.
\end{align}
\end{subequations}
The Cauchy-Schwarz inequality leads to
\begin{align}
    \left(\sum_{I=1}^M\avg{S_x^{I}}_{\rho_I}\right)^2\leq\sum_{I=1}^M\var{[S_z^{I}]}_{\rho_I}\sum_{I=1}^M\frac{\avg{S_x^{I}}_{\rho_I}^2}{\var{[S_z^{I}]}_{\rho_I}},
\end{align}
and we obtain the subadditivity
\begin{align}\label{eq:subadditivity}
    (2S)\xi^{-2}_{\rho_1\otimes\cdots\otimes\rho_M}\leq \sum_{I=1}^M (2S_I)\xi^{-2}_{\rho_I}.
\end{align}

\subsection{Limits on spin squeezing for $p$-producible states}
Consider a separable state of the form $\rho=\sum_{\gamma}p_{\gamma}\rho_{\gamma}^{1}\otimes\cdots\otimes\rho_{\gamma}^{M}$, where the $\rho_{\gamma}^{I}$ are density matrices of spin $S_I$-subsystems with $\sum_{I=1}^MS_I=S$. We obtain
\begin{subequations}
\begin{align}
    (2S)\xi^{-2}_{\rho}&\leq \sum_{\gamma}p_{\gamma}(2S)\xi^{-2}_{\rho_{\gamma}^{1}\otimes\cdots\otimes\rho_{\gamma}^{M}}&\text{using Eq.~(\ref{eq:convexity})}\\
    &\leq\sum_{\gamma}p_{\gamma}\sum_{I=1}^M(2S_I)\xi^{-2}_{\rho_{\gamma}^{I}}&\text{using Eq.~(\ref{eq:subadditivity})}\\
    &\leq\sum_{\gamma}p_{\gamma}\sum_{I=1}^M(2S_I)(S_I+1)&\text{using Eq.~(\ref{eq:minxi})}\\
    &=2\sum_{I=1}^MS_I^2+2S&.\label{eq:xim2bound}
\end{align}
\end{subequations}
Now we assume that the state $\rho$ is a $p$-producible state of $N$ spin-$1/2$ particles ($S=N/2$), \ie that each of its subsystems contains at most $p$ spin-$1/2$ particles. This sets the upper limit $S_I\leq p/2$ on the maximum spin length of each subsystem. Under this constraint, the function $\sum_{I=1}^MS_I^2$ is maximized by creating the largest possible number of $N_p=\lfloor N/p\rfloor$ groups of the maximal size $p$, and a single group of size $r=N-pN_p$ containing the remaining particles. This yields
\begin{align}
    \sum_{I=1}^MS_I^2\leq N_p \left(\frac{p}{2}\right)^2 +\frac{1}{4} r^2.
\end{align}
Inserting this into Eq.~(\ref{eq:xim2bound}) leads to the following condition for arbitrary $p$-producible $N$-qubit states:
\begin{align}
    N\xi^{-2}&\leq N_p \frac{p^2}{2} +\frac{r^2}{2} +N.
\end{align}
We may equivalently write this condition as
\begin{align}\label{eq:tighterboundWineland}
    \xi^{2}&\geq \frac{N}{N_p \frac{p^2}{2} +\frac{r^2}{2}+N}.
\end{align}
Note that, whenever $N/p$ is an integer, we have $N_p=N/p$, and thus $r=0$, and the bound simplifies to
\begin{align}
    \xi^{2}&\geq \frac{1}{1+p/2}.
\end{align}
According to Eq.~(\ref{eq:minxi}) this corresponds to the limit on spin squeezing for the largest entangled subsystem with spin $S=p/2$.

\section{Relation between the spin-squeezing entanglement witnesses of Wineland \textit{et al.}~\cite{Wineland} and S\o{}rensen-M\o{}lmer~\cite{SorensenMolmer}}\label{app:WSM}

The Wineland \textit{et al.} spin-squeezing coefficient~\cite{Wineland} expresses the ratio mean spin length and minimal variance in an orthogonal direction. This ratio has a clear metrological interpretation and, as we have discussed in Appendix~\ref{app:tighterbound}, it can be related to multiparticle entanglement. An alternative approach has been proposed by S\o{}rensen and M\o{}lmer~\cite{SorensenMolmer}, who use the combined information of mean spin length and minimal variance (beyond only their ratio) to derive limits on multiparticle entanglement.

Here, we demonstrate that the entanglement witness that is given by the Wineland \textit{et al.} spin-squeezing coefficient can be recovered from the approach of S\o{}rensen and M\o{}lmer in the limit of vanishing mean spin length, which corresponds to the scenario where their criterion is least effective.

\subsection{Properties of the functions $F_S[x]$}\label{app:FJ}

We first review and generalize the approach of S\o{}rensen and M\o{}lmer~\cite{SorensenMolmer}. We define $F_S$ as the minimum variance of $S_z$ divided by $S$ for a spin-$S$ system as a function of $\langle S_x\rangle$, \ie
\begin{equation}\label{FSboundVar}
    \dfrac{\var\left[ S_z \right]}{S} \geq F_S\left[ \dfrac{\langle S_x \rangle}{S} \right],
\end{equation}
holds for all states and can be saturated. A graphical illustration of these bounds is given in Fig.~\eqref{fig:FS} and in Ref.~\cite{SorensenMolmer}.

\begin{figure}[h!]
  \centering
\includegraphics[width=0.4\textwidth]{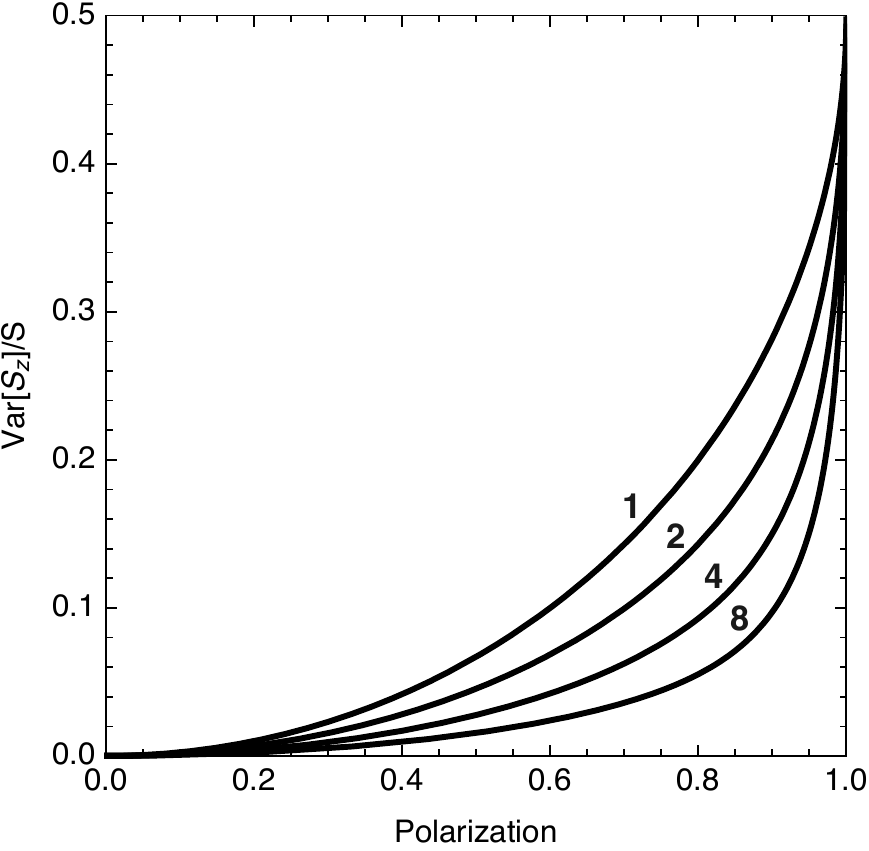}
  \caption{The S{\o}rensen-M{\o}lmer functions $F_S[x]$, Eq.~\eqref{FSboundVar}, for different $S$ as a function of the polarization $x=\avg{S_x}/S$ \cite{SorensenMolmer}.}
    \label{fig:FS}
\end{figure}

Here $F_S[x]$ with $x\in [-1,1]$, is a function with the following properties (we assume $S$ integer, see Ref.~\cite{SorensenMolmer} for a discussion of non-integer cases):
\begin{itemize}
    \item 0). $F_S[x]$ is symmetric in $x$, \ie $F_S[-x] = F_S[x]$, with $F_S[0]=0$ and $F[\pm 1]=1/2$.
    \item 1). $F_S[x]$ is convex in $x$, \ie $p F_S[x_1] + (1-p) F_S[x_2] \geq F_S[p x_1 + (1-p) x_2]$ for $0\leq p\leq 1$.
    \item 2). For $x\in[0,1]$, $F_S[x]$ is strictly increasing, \ie for $x_1<x_2$, we have $F_S[x_1]<F_S[x_2]$. 
    \item 3). For $S_1 < S_2$ we have $F_{S_1}[x] > F_{S_2}[x]$.
    \item 4). The $F_S[x]$ satisfy $\lim_{x\rightarrow 0} \dfrac{F_S[x]}{x^2} = \dfrac{1}{2+2S} + O(x^2)$. 
    \item 5). For $x\geq 0$ and $\lambda>1$ it holds
    $F_S[\lambda x]\geq \lambda F_S[x]$.
\end{itemize}
The symmetry of property 0) follows directly from the definition Eq.~\eqref{FSboundVar}, where the $x$-axis can always be chosen such that $\langle S_x\rangle\geq 0$. The value at $x=0$ follows from property 4), while the value at $x=1$ is attained by a spin-coherent state, with variance $\var[S_z]=S/2$. Properties 1) and 2) are proven in Ref.~\cite{SorensenMolmer}. Property 3) can be proven for large $S$ using the analytical expression~\cite{SorensenMolmer}
\begin{equation}\label{eq:MSanFS}
    F_S[x] = \dfrac{1}{2}\left(1 + S(1-x^2) - \sqrt{(1-x^2)((1+S)^2 - S^2 x^2)} \right) \;,
\end{equation}
valid for $S\gg 1$, while for smaller $S$, property 3) is confirmed numerically. Property 4) can be proven using the results Eqs.~(\ref{eq:pertSx},\ref{eq:pertSz2}) obtained from perturbation theory (note that the limit $x=\avg{S_x}/S \rightarrow 0$ corresponds to the limit $\lambda\rightarrow 0$). In this case, $\xi^2 / 2 \geq F_S[x]/x^2$, with equality when Eq.~\eqref{FSboundVar} is saturated and we obtain from Eq.~\eqref{eq:minxi} the desired limit. Note that if we were to take the limit using the expression Eq.~\eqref{eq:MSanFS}, we would have obtained $1/(4+4S)$. The factor two difference is attributed to the fact that Eq.~\eqref{eq:MSanFS} is an approximation that differs from the true bound by a factor two in the limit of small $x$, as mentioned in Ref.~\cite{SorensenMolmer}. Finally, to prove property 5) we first use that the convexity of $F_S[x]$ is equivalent to the condition
\begin{align}\label{eq:convexity}
    F_S[x]\geq F'_S[x_0](x-x_0)+F_S[x_0]
\end{align}
for all $x,x_0\in[0,1]$. Using $F_S[0]=0$, we obtain at $x=0$
\begin{align}\label{eq:convF0}
    F'_S[x_0]x_0\geq F_S[x_0].
\end{align}
At $x=\lambda x_0$ with $\lambda>1$, we can rewrite~(\ref{eq:convexity}) as
\begin{align}
    F_S[\lambda x_0]&\geq F'_S[x_0](\lambda-1)x_0+F_S[x_0]\notag\\
    &= (\lambda-1)(F'_S[x_0]x_0-F_S[x_0])+\lambda F_S[x_0]\notag\\
    &\geq \lambda F_S[x_0],
\end{align}
where in the last step, we used~(\ref{eq:convF0}).

\subsection{Generalization of the S\o{}rensen-M\o{}lmer bound to non-integer $N/p$}\label{app:SM}
S\o{}rensen and M\o{}lmer proved their criterion for a decomposition of the total spin $S$ into $N/p$ subgroups of size $p$, assuming that $N/p$ is integer. Here, we generalize their result by considering a separation of the total system into as many groups as possible of maximal size $p$ plus a remaining group. More precisely, call $N_p=\lfloor N/p \rfloor$ the number of partitions into groups of $p$ particles. Each group has spin $S_p=p/2$ (for spin-$1/2$ particles). If $N/p$ is not an integer, there will be an additional group of $r = N - p N_p$ particles, labeled $r$(est), with spin $S_r=r/2$. We obtain
\begin{align}
    \var\left[ S_z \right] &\geq \sum_{\gamma} p_{\gamma} \left( \sum_{i=1}^{N_p} \var\left[ S_z^{(i)} \right]_{\gamma} + \var\left[ S_z^{(r)} \right]_{\gamma} \right)&\text{concavity of the variance}\nonumber\\
    &\geq \sum_{\gamma} p_{\gamma} \left( \sum_{i=1}^{N_p} S_p \, F_{S_p}\left[ \dfrac{\langle S_x^{(i)} \rangle_{\gamma} }{S_p} \right] + S_r \, F_{S_r}\left[ \dfrac{\langle S_x^{(r)} \rangle_{\gamma} }{S_r} \right] \right) &\text{using Eq.~(\ref{FSboundVar})}\nonumber\\
    &\geq \sum_{\gamma} p_{\gamma} \left( (S_p N_p) \, F_{S_p}\left[ \dfrac{1}{S_p N_p} \sum_{i=1}^{N_p} \langle S_x^{(i)} \rangle_{\gamma} \right] + S_r \, F_{S_r}\left[ \dfrac{\langle S_x^{(r)} \rangle_{\gamma} }{S_r} \right] \right) &\text{using property 1)}\nonumber\\
    &\geq (S_p N_p) \, F_{S_p}\left[ \dfrac{1}{S_p N_p} \sum_{i=1}^{N_p} \sum_{\gamma} p_{\gamma} \langle S_x^{(i)} \rangle_{\gamma} \right] + S_r \, F_{S_r}\left[ \sum_{\gamma} p_{\gamma} \dfrac{\langle S_x^{(r)} \rangle_{\gamma} }{S_r} \right] &\text{using property 1)} \nonumber\\
    &= (S_p N_p) \, F_{S_p}\left[ \dfrac{1}{S_p N_p} \sum_{i=1}^{N_p} \langle S_x^{(i)} \rangle \right] + S_r \, F_{S_r}\left[ \dfrac{\langle S_x^{(r)} \rangle }{S_r} \right].\label{eq:tighterboundSM}
\end{align}

This bound is the tightest formulation of the multipartite entanglement criterion first proposed by S\o{}rensen and M\o{}lmer in Ref.~\cite{SorensenMolmer}. It can be further simplified under the assumption of symmetric spin states. If property (ii) (see Sec.~\ref{sec:symmetry} in the main text) is granted, we may write $\langle S_x^{(i)} \rangle=p\langle S_x\rangle/N$ and $\langle S_x^{(r)}\rangle = r\langle S_x\rangle/N$. This yields
\begin{align}
    \var\left[ S_z \right]&\geq   (S_p N_p) \, F_{S_p}\left[ \dfrac{1}{S_p N_p} N_p \frac{p}{N} \langle S_x\rangle\right] + S_r \, F_{S_r}\left[ \frac{1}{S_r}\dfrac{r}{N}\langle S_x \rangle  \right] \nonumber\\
    &=(S_p N_p) \, F_{S_p}\left[\frac{\langle S_x\rangle}{S}\right] + S_r \, F_{S_r}\left[\frac{\langle S_x\rangle}{S}\right],\label{MSgeneralizedApp}
\end{align}
where $S=N/2$ is the total spin. 

By explicitly considering a subgroup of size $r$ when the distribution of $N$ particles into subgroups of size $p$ cannot account for all particles, this bound is stronger than the well-known bounds derived in~\cite{SorensenMolmer}. To see this, notice that $S_r<S_p$ and by virtue of property 3) we can thus derive the weaker bound
\begin{align}
    \var\left[ S_z \right]&\geq
    (S_p N_p) \, F_{S_p}\left[\frac{\langle S_x\rangle}{S}\right] + S_r \, F_{S_p}\left[\frac{\langle S_x\rangle}{S}\right]\nonumber\\
    &= S \, F_{S_p}\left[\frac{\langle S_x\rangle}{S}\right],
\end{align}
with $S=S_p N_p+S_r$.
This is the standard formulation of the S{\o}rensen-M{\o}lmer multiparticle entanglement criterion~\cite{SorensenMolmer} and we observe that the condition~(\ref{MSgeneralizedApp}) is indeed stronger. If $N_p$ is an integer, both criteria coincide, since $S_r=0$ and $S_pN_p=S$. 

\subsection{Relation between different spin-squeezing multiparticle entanglement criteria}\label{app:MSWINE}
Since the functions $F_S[x]$ are strictly increasing, the entanglement criterion~(\ref{MS}) detects the largest number of entangled states when $\avg{S_x}$ is large. 
If, however, the value of $\avg{S_x}$ is not known separately from $\var{[S_z]}/\avg{S_x}^2$, we must assume the `worst-case scenario', which consequently is given in the limit of $\avg{S_x}\rightarrow 0$. Here, we show that in this limit, the S\o{}rensen-M\o{}lmer approach to witnessing entanglement becomes equivalent to the bound~(\ref{WkprodTight}) on the Wineland \textit{et al.} spin-squeezing coefficient. We first demonstrate this correspondence for the simple bounds, \ie Eqs.~(\ref{WkprodTight}) and~(\ref{MS}) and then generalize our result to the tighter expressions~(\ref{eq:tighterboundWineland}) and~(\ref{eq:tighterboundSM}).

Note that the condition~(\ref{MS}) for states with at most $p$-partite entanglement can be equivalently stated as
\begin{align}
    \frac{N\var\left[ S_z \right]}{\avg{S_x}^2}&\geq
    2\, \frac{F_{S_p}\left[\frac{\langle S_x\rangle}{S}\right]}{\left(\frac{\avg{S_x}}{S}\right)^2}.
\end{align}
Making use of properties 2 and 4, we further obtain that
\begin{align}
    2\, \frac{F_{S_p}\left[\frac{\langle S_x\rangle}{S}\right]}{\left(\frac{\avg{S_x}}{S}\right)^2}&\geq
    \frac{1}{\frac{p}{2}+1}.
\end{align}
Combining these two bounds, we can derive the multiparticle spin squeezing condition~(\ref{WkprodTight}) in the `worst-case' limit of the approach of Eq.~(\ref{MS}).

Let us now demonstrate that the same correspondence holds for the most general formulation of the respective criteria for non-integer $N/p$. Using properties~2 and~4 in Eq.~(\ref{eq:tighterboundSM}) implies that in the limits $\langle S_x^{(i)} \rangle\to 0$ and $\langle S_x^{(r)} \rangle\to 0$ the following condition holds for states with no more than $p$-partite entanglement:
\begin{align}
    \var\left[ S_z \right] 
    &\geq \frac{\left(\sum_{i=1}^{N_p} \langle S_x^{(i)} \rangle\right)^2}{N_p\frac{p^2}{2}+pN_p}  +  \frac{ \langle S_x^{(r)} \rangle^2}{\frac{r^2}{2}+r}.\label{eq:intermediate}
\end{align}
We now demonstrate that the following statement holds for arbitrary $A, B, a,b\in \mathbb{R}$ with $a,b>0$:
\begin{align}\label{eq:ABbound}
    \frac{A^2}{a}+\frac{B^2}{b}\geq \frac{(A+B)^2}{a+b}.
\end{align}
This relation follows immediately from
\begin{align}
   0\leq \frac{(A b - a B)^2}{a b (a + b)}= \frac{A^2}{a}+\frac{B^2}{b}- \frac{(A+B)^2}{a+b}.
\end{align}
Using Eq.~(\ref{eq:ABbound}) with $A=\sum_{i=1}^{N_p} \langle S_x^{(i)} \rangle$, $B=\langle S_x^{(r)} \rangle$, $a=N_p\frac{p^2}{2}+pN_p$, and $b=\frac{r^2}{2}+r$, we obtain from Eq.~(\ref{eq:intermediate}) that
\begin{align}\label{eq:intermediate2}
    \var\left[ S_z \right] 
    &\geq \frac{\avg{S_x}^2}{N_p\frac{p^2}{2}+\frac{r^2}{2}+N},
\end{align}
where we have used that $A+B=\sum_{i=1}^{N_p} \langle S_x^{(i)}\rangle+\langle S_x^{(r)}\rangle=\avg{S_x}$ and $pN_p+r=N$. Multiplying both sides of~(\ref{eq:intermediate2}) by $N$ and dividing by $\avg{S_x}^2$, we recover the condition~(\ref{eq:tighterboundWineland}). This generalizes the correspondence between the two approaches to the stronger conditions, valid in the case of non-integer $N/p$.

\subsection{Detecting genuine $p$-partite entanglement}\label{app:genpMS}
Detecting genuine $(p+1)$-partite entanglement (recall Appendix~\ref{app:ent}) requires to exclude not only each $p$-producible model individually but also all convex combinations of the kind 
\begin{align}\label{eq:gpprod}
\rho=\sum_{\Lambda\in\mathfrak{L}_{p-\mathrm{prod}}}P_{\Lambda}\rho_{\Lambda-\mathrm{sep}} \;,
\end{align}
where each of the $\rho_{\Lambda-\rm{sep}}$ is of the form of Eq.~(\ref{eq:Lsep}).
From the convexity of the spin-squeezing coefficient $\xi^{-2}_{\rho}$ [see Eq.~(\ref{eq:convexity})] we can immediately conclude that the bounds for $p$-producible states [see, \eg Eq.~(\ref{WkprodTightNp})] also hold for arbitrary linear combinations of $p$-producible states, since the bounds are state independent.

This is not the case for the tighter bounds proposed by S\o{}rensen and M\o{}lmer, Eq.~\eqref{MS}: These depend on the polarization which could be in principle different in each of the states $\rho_{\Lambda-\mathrm{sep}}$. The convexity property of the functions $F_S$ nevertheless allows us to interpret Eq.~\eqref{MS} as a criterion for genuine multipartite entanglement~\cite{ReidHeDrummundFrontiers2011}.

We first establish the following property: Assume $S_2 > S_1$, from the properties of the $F_S$ function we can write
\begin{subequations}\label{eq:derivedFprop}
\begin{align}
S_1 F_{S_1}[x_1] + S_2 F_{S_2}[x_2] &\geq S_1 F_{S_2}[x_1] + S_2 F_{S_2}[x_2] &\;\text{using 3)} \\
&= (S_1+S_2)\left( \dfrac{S_1}{S_1+S_2} F_{S_2}[x_1] + \dfrac{S_2}{S_1+S_2} F_{S_2}[x_2] \right) \\
&\geq (S_1+S_2) F_{S_2}\left[\dfrac{S_1 x_1 + S_2 x_2}{S_1+S_2}\right] \;. &\;\text{using 1)}
\end{align}
\end{subequations}

Consider the state Eq.~(\ref{eq:gpprod}) that is a convex combination of arbitrary $p$-producible models. Recall that each decomposition of the form of Eq.~(\ref{eq:Lsep}) depends on $\Lambda$, even though we do not make this dependence explicit below to simplify our notation. We further denote the total spin of subsystem $\mathcal{A}_q$ as $S_q=|\mathcal{A}_q|/2$. Using the concavity of the variance, we obtain
\begin{align}
    \var\left[ S_z \right] &\geq \sum_{\Lambda}\sum_{\gamma} P_{\Lambda}p_{\gamma} \left( \sum_{q=1}^{k} \var\left[ S_z^{(k)} \right]_{\gamma} \right)\nonumber\\
    &\geq \sum_{\Lambda}\sum_{\gamma} P_{\Lambda}p_{\gamma} \left( \sum_{q=1}^{k} S_q \, F_{S_q}\left[ \dfrac{\langle S_x^{(q)} \rangle_{\gamma} }{S_q} \right]\right) \;.
\end{align}
Noticing that the largest possible value of all the $S_q$ is determined by $S_p=p/2$, we can apply property~(\ref{eq:derivedFprop}) successively to bound the sum over $q$, which gives
\begin{align}
    \var\left[ S_z \right] &\geq \sum_{\Lambda}\sum_{\gamma} P_{\Lambda}p_{\gamma} S \, F_{S_p}\left[ \dfrac{\langle S_x \rangle_{\gamma} }{S} \right] \;,
\end{align}
where $S=\sum_{q=1}^kS_q$ and $\langle S_x \rangle_{\gamma}=\sum_{q=1}^k\langle S_x^{(q)} \rangle_{\gamma}$. Finally, using convexity of the $F_S$ functions [property 1)], we obtain the bound
\begin{align}
    \var\left[ S_z \right] &\geq S \, F_{S_p}\left[\sum_{\Lambda}\sum_{\gamma} P_{\Lambda}p_{\gamma}  \dfrac{\langle S_x \rangle_{\gamma} }{S} \right]\nonumber\\
    &=S \, F_{S_p}\left[\dfrac{\langle S_x \rangle}{S} \right] \;.
\end{align}
Therefore, finding the maximum integer $p$ for which this inequality is violated allows us to conclude that the state of the system is genuine $(p+1)$-partite entangled.

\subsection{EPR steering criterion based on the S\o{}rensen-M\o{}lmer bounds Eq.~\eqref{MS}}\label{app:SMepr}

Consider a fixed bi-partition of the system into $N_A$, $N_B=N-N_A$ particles. Steering of party $B$ by $A$ can be detected from the following criterion based on the S\o{}rensen-M\o{}lmer bounds Eq.~\eqref{MS}.
We first use the concavity of the variance to obtain
\begin{subequations}
\begin{align}\label{eq:EPRMS}
    \var\left[ S_z \right] &\geq \sum_{\gamma} p_{\gamma} \left( \var\left[ S_z^{A} \right]_{\gamma} + \var\left[ S_z^{B} \right]_{\gamma} \right) \\
    &\geq \sum_{\gamma} p_{\gamma} \var\left[ S_z^{B} \right]_{\gamma} \\
    &\geq \sum_{\gamma} p_{\gamma} 
    S_B \, F_{S_B}\left[ \dfrac{\langle S_x^{B} \rangle_{\gamma} }{S_B} \right]  \\
    &\geq 
    S_B \, F_{S_B}\left[ \sum_{\gamma} p_{\gamma} \dfrac{\langle S_x^{B} \rangle_{\gamma} }{S_B} \right] \\
    &= S_B \, F_{S_B}\left[ \dfrac{\langle S_x^{B} \rangle }{S_B} \right] \;.
\end{align}
\end{subequations}
In the second step, we used the fact that in absence of a local quantum description of system $A$, we can only assume that $\var[S_z^A]\geq 0$, and the following steps follow from the properties of the $F_S$ functions.

In Ref.~\cite{ReidHeDrummundFrontiers2011}, it was shown that steering is detected by a violation of the bound
\begin{align}\label{eq:ReidCondEPRMS}
    \var\left[ S_z \right] \geq S \, F_{S_B}\left[ \frac{\langle S_x^{B} \rangle }{S} \right].
\end{align}
We now show that our condition~(\ref{eq:EPRMS}) implies Eq.~(\ref{eq:ReidCondEPRMS}) and is therefore a stronger steering witness. To see this, we denote $\lambda=S/S_B>1$ and write $S_B \, F_{S_B}[ \dfrac{\langle S_x^{B} \rangle }{S_B}]=(S/\lambda)F_{S_B}[\lambda \dfrac{\langle S_x^{B} \rangle }{S}]\geq SF_{S_B}[\dfrac{\langle S_x^{B} \rangle }{S}]$, where in the last step we used the superlinear scaling of the $F_S[x]$, property 5) in Appendix~\ref{app:FJ}.

\newpage

\end{widetext}

\end{document}